\documentclass[aip,apl,preprint]{revtex4-1}
\usepackage{graphicx}
\usepackage{dcolumn}
\usepackage{bm}
\usepackage{subfigure}
\usepackage{color}
\usepackage{float}
\usepackage{amsmath}
\usepackage{gensymb}

\begin{document}

\title{{Full-field hard x-ray microscopy with interdigitated silicon lenses}}
\date{\today}

\author{Hugh \surname{Simons}}
\email{husimo@fysik.dtu.dk}
\affiliation{Department of Physics, Technical University of Denmark, Building 307, Kgs. Lyngby DK-2800, Denmark}
\affiliation{European Synchrotron Radiation Facility, Grenoble 38000, France}

\author{Frederik \surname{St\"ohr}}
\affiliation{DANCHIP, Technical University of Denmark, Building 347, Kgs. Lyngby DK-2800, Denmark}

\author{Jonas \surname{Michael-Lindhard}}
\affiliation{DANCHIP, Technical University of Denmark, Building 347, Kgs. Lyngby DK-2800, Denmark}

\author{Flemming \surname{Jensen}}
\affiliation{DANCHIP, Technical University of Denmark, Building 347, Kgs. Lyngby DK-2800, Denmark}

\author{Ole \surname{Hansen}}
\affiliation{DTU Nanotech, Technical University of Denmark, Building 345E, Kgs. Lyngby DK-2800, Denmark}
\affiliation{CINF, Technical University of Denmark, Building 345E, Kgs. Lyngby DK-2800, Denmark}

\author{Carsten \surname{Detlefs}}
\affiliation{European Synchrotron Radiation Facility, Grenoble 38000, France}

\author{Henning Friis \surname{Poulsen}}
\affiliation{Department of Physics, Technical University of Denmark, Building 307, Kgs. Lyngby DK-2800, Denmark}

\begin{abstract}
Full-field x-ray microscopy using x-ray objectives has become a mainstay of the biological and materials sciences. However, the inefficiency of existing objectives at x-ray energies above 15 keV has limited the technique to weakly absorbing or two-dimensional (2D) samples. Here, we show that significant gains in numerical aperture and spatial resolution may be possible at hard x-ray energies by using silicon-based optics comprising `interdigitated' refractive silicon lenslets that alternate their focus between the horizontal and vertical directions. By capitalizing on the nano-manufacturing processes available to silicon, we show that it is possible to overcome the inherent inefficiencies of silicon-based optics and interdigitated geometries. As a proof-of-concept of Si-based interdigitated objectives, we demonstrate a prototype interdigitated lens with a resolution of $\approx255$ nm at 17 keV. 
\end{abstract}



\maketitle
 
X-ray microscopy (XRM) is an established family of techniques for imaging embedded and structurally complex specimens with sub-$\mu$m resolution. The ability to ‘look’ inside dense matter has provided crucial insight into phenomena such as ferromagnetic domains\cite{Fischer1998}, nano-scale strain \cite{Hruszkewycz2012} and compositional inhomogeneity \cite{McNeill2007}. The techniques can be broadly categorized into \emph{scanning-} \cite{Thibault2008}, \emph{projection-} \cite{McNulty1992} and \emph{objective-}based\cite{Lengeler1999} approaches, which offer different compromises of spatial resolution, acquisition time and sensitivity. Full-field imaging with an objective is particularly relevant to materials and geological sciences, as its efficiency and modalities (e.g. dark-field \cite{Simons2015}) enable real-time imaging of complex processes \cite{VanWaeyenberge2006}. Performing full-field XRM at hard x-ray energies ($>$15 keV) would then open a new door to dynamic, three-dimensional (3D) multi-scale studies of denser and more complex samples. However, current x-ray objectives tend to be aberrated or inefficient in the hard x-ray regime, limiting spatial and temporal resolution. 

In this letter, we show that Si-based 2D compound refractive lenses (CRLs) are a viable approach for improving the numerical aperture (NA) and efficiency of full-field XRM objectives at hard energies. Specifically, we show theoretically that sufficiently miniaturized Si CRLs can outperform the current state-of-the-art at hard energies, and validate this prediction based on the performance of a prototype objective at 17 keV.
 
\begin{figure}
\includegraphics[width=8.5cm]{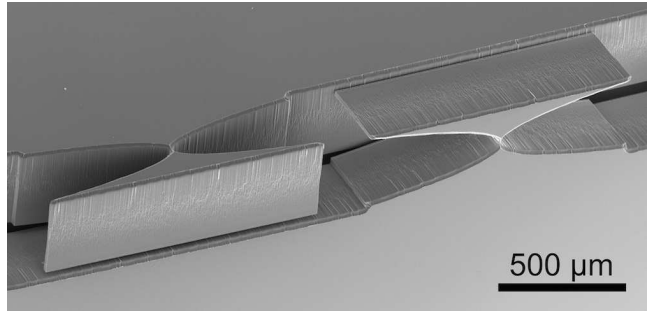}
\caption{SEM image of Si interdigitated lenslets within the prototype objective lens tested in this work. The small apex radii of curvature accessable to Si-based lenslets enables shorter focal lengths and higher NAs than existing 2D objectives}\label{fig1}
\end{figure}

Various x-ray imaging optics have been proposed for use at hard energies: reflective optics (multilayer mirrors \cite{Underwood1986}) are efficient but expensive and delicate; diffractive optics (Fresnel zone plates \cite{Niemann1976}) have large NAs but are inefficient, while refractive optics (compound lenses \cite{Snigirev1996} or prisms \cite{Marschall2014}) can be efficient but are prone to aberration and small NAs. The approach demonstrated here can overcome some of the current limitations of refractive optics at hard energies by utilizing a miniaturized, 2D ‘interdigitated’ configuration of planar silicon lenses (Fig. \ref{fig1}).

The ideal lenslet geometry and configuration of a CRL-based XRM can be determined by optimizing the NA. From ref. \onlinecite{Lengeler1999}, the NA of an XRM using an ideal CRL of focal length $f$, comprising $N$ identical lenslets with apex radius of curvature $R$, linear attenuation coefficient $\mu$ and refractive decrement $\delta$ follows the relationship:

\begin{equation}
\mathrm{NA} \propto \left(\frac{\mathcal{M}}{\mathcal{M}+1}\right) \sqrt{\frac{\delta}{\mu}} \sqrt{\frac{1}{f}}
= \left(\frac{\mathcal{M}}{\mathcal{M}+1}\right) \sqrt{\frac{N \delta^2}{\mu R}}\label{eq1}
\end{equation}

Eq. \ref{eq1} implies that for a given magnification $\mathcal{M}$, the NA is greatest in CRLs with a short focal length and lenslets of a low-Z material. Many CRLs therefore utilize lenslets produced by indenting the parabolic lens profile into polycrystalline Be or Al\cite{Lengeler1999}. While these materials have favourable values of $\delta$ and $\mu$ in the hard x-ray regime, the indenting process is expensive and limited to large $R$ due to material (grain structure, porosity, plasticity) and processing (tool shape, concentricity) issues. At hard energies, this necessitates the use of many lenslets at significant cost to achieve a short $f$ and thus large NA. Polymer CRLs avoid these drawbacks, but are susceptible to radiation damage when flux and energy are high (e.g. in bright-field XRM) \cite{Snigireva2004}. However, eq. \ref{eq1} also indicates an alternative route to large NAs by using lenses of inferior refractive medium but with drastically miniaturized dimensions (i.e. small $R$ and lenslet thickness $T$).

We therefore propose Si-based objectives produced using the same nano-manufacturing methods as for nano-focusing 1D CRLs, which are both cost-effective and capable of apex radii as small as 1 $\mu$m \cite{Schroer2003}. 2D focusing can be achieved by `interdigitating' the 1D CRLs such that horizontal and vertical lenslets alternate\cite{Somogyi2011} as shown in Fig. \ref{fig1}. This reduces astigmatic aberration compared to sequential chips and potentially enables lens-by-lens optimization of the NA through aberration-corrected \cite{Marschall2014}, adiabatic \cite{Schroer2005} and kinoform geometries \cite{Evans-Lutterodt2007}. Ultimately, we show theoretically that the technical benefits of utilising Si as a refractive medium overcomes its inherent disadvantages in terms of refractive performance to yield higher performing lenses at hard energies.

The formalism\cite{supp_mater} uses the ray-transfer matrix approach\cite{Protopopov1998,Pantell2003,Poulsen2014} to describe the cumulative effects of the individual lenslets in the CRL and predict the aberration induced by misalignment. At its core is the general expression for the transmission function $I/I_0$. In the 1D case of axisymmetric imaging lenses, $I/I_0$ describes the attenuation of a ray originating from a Gaussian source with radial position $r_s$ and angle $w_s$ by a CRL comprising $N$ identical parabolic lenslets of web thickness $T_0$ and linear attenuation coefficient $\mu$:

\begin{equation}
	\begin{split}
		\frac{I}{I_0}(r_s,w_s) & \propto \exp{(-NT_0\mu)} \times \exp{\left(\frac{-r_s^2}{2\sigma_v^2}\right)} \\
		& \times\exp{\left[\frac{-(w_s-\gamma r_s)^2}{2\sigma_a^2}\right]}\label{eq2}
	\end{split}
\end{equation}

\begin{figure}
\includegraphics[width=8.5cm]{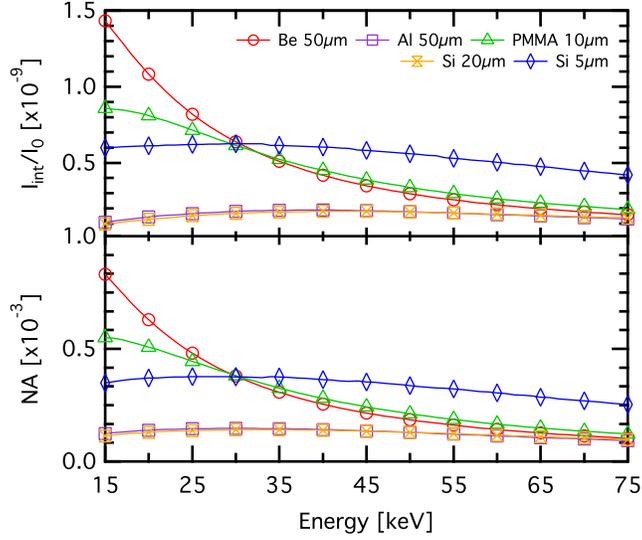}
\caption{NA and total integrated image intensity ($I_{tot}/I_0$) of axisymmetric Be and Al lenses compared to interdigitated polymer and Si lenses in the hard x-ray regime. The number of lenses (and hence focal length) is optimized for each CRL and energy to produce the best compromise of NA and $I_{tot}/I_0$. Details of the optimization routine and the CRL geometries are given in ref \onlinecite{supp_mater}}\label{fig2}
\end{figure}

Eq. \ref{eq2} is a product of three exponential terms: a constant absorption factor, a Gaussian with standard deviation $\sigma_v$ describing the vignetting at the sample plane (i.e. reduction in image brightness towards the periphery) and another Gaussian describing the angular acceptance of the lens whose standard deviation $\sigma_a$ is half the NA. We can therefore optimize the NA in terms of $N$ and $R$ (and consequently the sample-objective distance $d_1$) for a given material, magnification and x-ray energy. This then enables the comparison of the best practically-achievable performance of CRLs of different materials geometries and configurations across the hard energy regime. Fig. \ref{fig2} shows the optimum NA and the total integrated image intensity from 15 to 75 keV for a selection of imaging objectives: Axisymmetric lenses of Be and Al (both $R$ = 50 $\mu$m), and interdigitated lenses of PMMA ($R$ = 10 $\mu$m), Si ($R$ = 20 $\mu$m and $R$ = 5 $\mu$m). The advantages of small-$R$ interdigitated Si CRLs are evident above $\approx$30-35 keV, where they can be more efficient than the PMMA, Be or Al CRLs.

We produced a prototype interdigitated lens comprising two chips of 20 1D Si lenslets with 20.5$\pm 0.016$ $\mu$m apex radii. The chips were produced through the standard contact UV-lithographic and deep reactive ion etching process \cite{Wu2010} to a depth of 350 $\mu$m, laser-cut from the wafer and assembled on a steel gauge block using a micromanipulator and an optical microscrope. The 2D lens was tested in a full-field XRM at beamline ID06 at the European Synchrotron using incoherent x-rays of energy 17 keV (wavelength $\lambda$ = 0.7293 {\AA}). The microscope had a magnification ratio of $\mathcal{M}$=11 and was intentionally diffraction-limited by the objective and oversampled by the CCD detector to reduce the contribution of the detector optics to the optical transfer function and resolution.

\begin{figure}
\includegraphics[width=8.5cm]{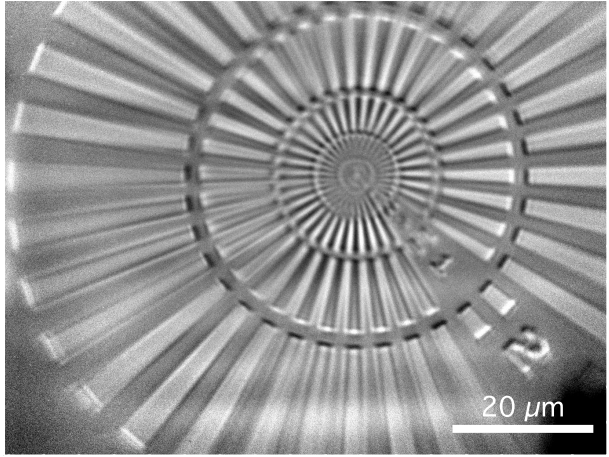}
\caption{Image of resolution chart at $\mathcal{M}$=11. The image has been post-processed by background subtraction and flat-field correction }\label{fig3}
\end{figure}

A direct and straightforward measure of the resolution and aberration of the microscope can be made by inspecting the image of an absorbing Siemens star with radial features ranging from 5 $\mu$m to 50 nm (Fig. \ref{fig3}). The image shows a rapid degradation of contrast between 500-200 nm as well as horizontal striations of varying intensity. We attribute the contrast variation across the image to a combination of partial coherence of the incident beam and small shape errors in the lenslets.

\begin{figure}
\includegraphics[width=8.5cm]{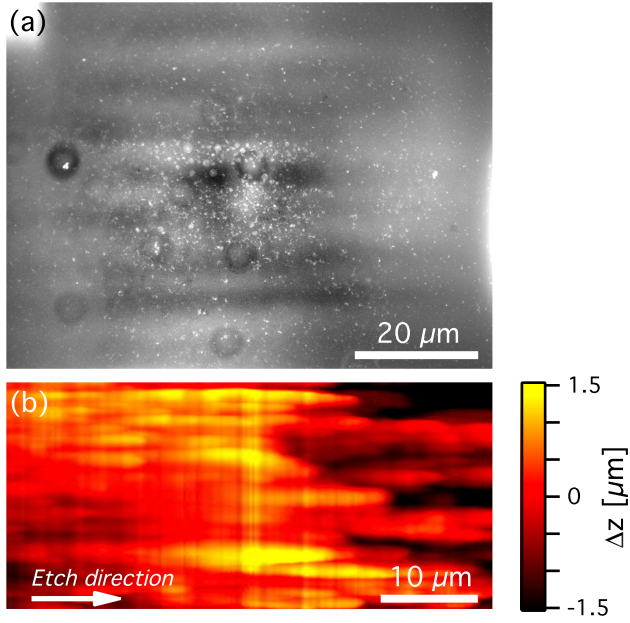}
\caption{Comparison of (a) heterogeniety in the flat-field residual and (b) surface topology of a single lens}\label{fig4}
\end{figure}

Imaging the flat wavefront (i.e. without the sample) yields a magnified intensity distribution related to the vignetting function in eq. \ref{eq2}. Fitting a 2D Gaussian to this distribution and inspecting its broadness and residual provides a means to compare the attenuation profile of the CRL to theoretical expectations and to characterise shape errors in lens profile. For this lens, the Gaussian intensity function had standard deviations of 19.06 $\mu$m and 18.53 $\mu$m in the horizontal and vertical directions respectively, which compare favourably with the theoretically-calculated values of 18.57 $\mu$m and 18.44 $\mu$m. The $\approx$2.8\% difference between the horizontal and vertical values suggest a slight astigmatism. The residual from this fit (Fig. \ref{fig4}(a)) shows diffuse horizontal striations across the field of view. The surface topography of a single lenslet measured by atomic force microscopy \cite{Stohr2015} shows ridges of similar location, aspect ratio and frequency, suggesting that the diffuse striations and irregular image contrast may originate from shape errors incurred during etching (Fig. \ref{fig4}(b)).

A more quantifiable measure of optical performance comes from the optical transfer function (OTF), $\mathcal{H}(f_X,f_Y)$, represented in a convolution with the intensity distribution at the sample $I/I_0(x_s,y_s)$ to result in the intensity distribution at the detector $I/I_0(x_d,y_d)$ \cite{Goodman1968} (where $\mathcal{F}^{-1}$ denotes the inverse Fourier transform):

\begin{equation}
I(x_d,y_d) = \mathcal{F}^{-1}|\mathcal{H}(f_X,f_Y)|^2 \otimes I(x_s,y_s)\label{eq3}
\end{equation}

For a non-aberrated axisymmetric CRL, $|\mathcal{H}|^2$ (the modulation transfer function, MTF) is a Gaussian function in the frequency ($f_R$) domain:

\begin{equation}
|\mathcal{H}(f_R)|^2 = \exp{\left(-\frac{f_R^2 \lambda^2}{4 \sigma_a ^2}\right)}\label{eq4}
\end{equation}

It then follows that by imaging a sample whose Fourier transform (FT) is known \emph{a priori}, we can determine the MTF and calculate $\sigma_a$ and thus NA. To this end, we measured a high-resolution image of a square grid of absorbing gold with band width $W = 5$ $\mu$m and period $P = 10$ $\mu$m. The FT of such a grid is well approximated by a series of 2D laplacian distributions with broadness $\sigma_L$. Thus, including $\mathcal{H}$ and a linear background: $B(f_R) = (b_0 + b_1 f_R)$, the spatial frequency distribution of the image, $G(f_R)$ is:

\begin{equation}
\begin{split}
G(f_R) & = B(f_R) + \mathcal{H}(f_R) \\ 
& \times \sum_{n=-N}^{N} \textrm{sinc} \left(\frac{2 \pi n W}{P}\right)\exp{\left({\frac{-|r_s-\frac{2 \pi n}{P}|}{\sigma_L}}\right)}\label{eq5}
\end{split}
\end{equation} 

\begin{figure}
\includegraphics[width=8.5cm]{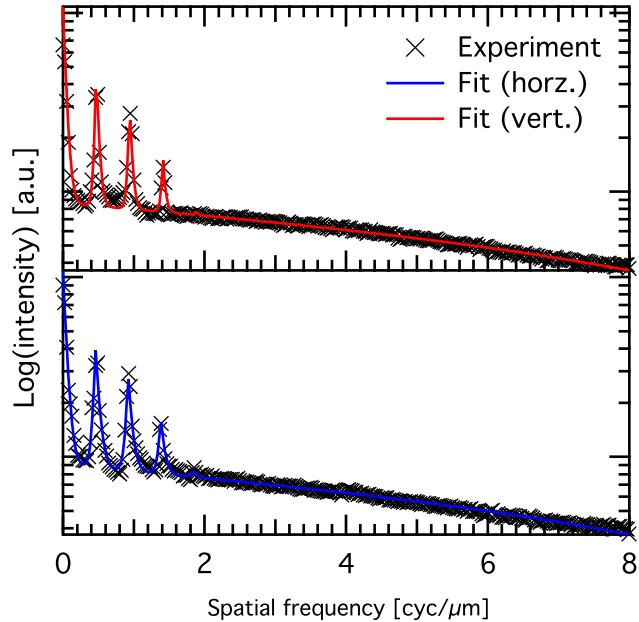}
\caption{Horizontal and vertical Fourier tranforms of the grid image and their respective fits with the theory in eq. \ref{eq5}}\label{fig5}
\end{figure}

Fig. \ref{fig5} shows the results from fitting the horizontal and vertical FTs of the grid pattern with Eq. \ref{eq5}. This yielded values for $\sigma_a$ of 5.94 $\mu$rad and 5.33 $\mu$rad in the vertical and horizontal directions, which compare favourably to the theoretical predictions of 5.39 $\mu$rad and 5.36 $\mu$rad. The cutoff frequency $f_0$ was solved numerically from the fit results \cite{Roth2015} and used to calculate resolution values of 230 and 280 nm for the vertical and horizontal directions. The difference between the $\sigma_a$ and $f_0$ values are also consistent with astigmatism, which was measured directly as the difference in focal position $\Delta f$ where $f_0$ is maximized in the horizontal and vertical directions. The 8 mm difference can be attributed to an angular misalignment (i.e. non-perpendicularity of the two chips) of $\approx 0.75 \degree$, which is related to the lenslet focal length $f$ and misalignment angle $\theta$ according to eq. \ref{eq6}:

\begin{equation}
\Delta f = f_x - f_y \simeq \frac{2f\sin{(2\theta)}}{N\cos^2{(2\theta)}} \approx 4\theta\frac{f}{N}\label{eq6}
\end{equation}

While small astigmatisms are typically trivial due to the large depth of field for CRL-based XRMs \cite{Lengeler1999}, we note that any predictable aberration could be remedied by the addition of corrective focusing elements \cite{Marschall2014}. This highlights an important advantage of interdigitated geometries in the ability to incorporate optimized and aberration-corrected geometries. One could maximise the NA beyond that theorized in Fig.~\ref{fig2} by optimizing a) refractive power, by progressively altering the lens profile through the CRL (i.e. adiabatic geometry \cite{Schroer2005}) and b) transmission, by adopting kinoform or Fresnel geometries \cite{Evans-Lutterodt2007}. We estimate that improvements of 1.5-2$\times$ to the NA and 2-4$\times$ to the transmission are within reach, constituting a significant step towards high-speed lenses for dynamic XRM. We also note that this technology could make even higher x-ray energies accessible through electroplated Ni lenses \cite{Nazmov2004}.

In conclusion, we have shown that due to their potential for miniaturization and short focal lengths, Si-based interdigitated optics can in principle outperform conventional, Be-based optics above 30 keV. To validate this prediction, we demonstrated a functioning 2D silicon objective for use in a full-field XRM at hard energies. The results are promising; showing acceptably low aberration and performance close to theoretical expectations. By harnessing the potential for miniaturization and aberration-correction of Si-based nano-focusing lenses, one could significantly improve the focusing power, transmission and NA at hard energies, potentially realizing hard-XRM with sub-100 nm resolution. Furthermore, the improved focal power available to these lenses would enable shorter imaging distances that are both more mechanically stable and practically achievable in modest synchrotron beamline hutches or laboratories. 

\begin{acknowledgements}

We thank the European Synchrotron for providing beamtime, Danish Fundamental Metrology for providing AFM use, and S.R. Ahl and J. Madsen for discussions. This work was supported by an ERC Advanced Grant; `d-TXM', and the Danish Instrumentation Centre, Danscatt. H.S. ackowledges support from the Danish Council for Independent Research individual postdoc grant, and O.H. acknowledges the Danish National Research Foundations' Centre for Individual Nanoparticle Functionality (DNR54).
\end{acknowledgements}

\bibliography{article.bib}

\end{document}